# Direct experimental measurement of phase-amplitude coupling in spin torque oscillators

L. Bianchini,[1,a)] S. Cornelissen,[2,3] Joo-Von Kim,[1,b)] T. Devolder,[1] W. van Roy,[2] L. Lagae,[2,4] and C. Chappert[1]

[1]*Institut d'Electronique Fondamentale, Univ. Paris-Sud, 91405 Orsay, France, and UMR 8622, CNRS, 91405 Orsay, France*
[2]*IMEC, Kapeldreef 75, 3001 Leuven, Belgium*
[3]*Electrical Engineering (ESAT), KU Leuven, Leuven, Belgium*
[4]*Department of Physics and Astronomy, KU Leuven, Leuven, Belgium*

We study spin-torque induced oscillations of MgO magnetic tunnel junctions in the time domain. By using the Hilbert transform on the time traces, we obtain for the first time a direct experimental measure of the coupling between the power and the phase fluctuations. We deduce the power restoration rate and we obtain low values for the coupling strength, which is consistent with the weak frequency dependence on the applied voltage. *© 2010 American Institute of Physics*. [doi:10.1063/1.3467043]

An electrical current flowing through a ferromagnetic layer is spin-polarized and can exert a torque on a second thin ferromagnetic layer's magnetization, as predicted by Slonczewski[1] and Berger.[2] In certain geometries and even for zero applied field, a steady state precession of the magnetization is obtained above a certain threshold when the torque applied by the injected current is large enough to compensate the natural Gilbert damping. In magnetic tunnel junctions (MTJ), the sustained precession of magnetization is converted into microwave voltage signal by the tunnel magnetoresistance effect. It is known that such spin-torque induced nano-oscillator (STO) is nonlinear,[3,4] i.e., its frequency depends on the oscillation power $p$.[5–9]

Although nonlinearity is an advantage for applications such as voltage-controlled nanoscale microwave oscillators, it is detrimental when a narrow linewidth is required. According to spin-wave theory, the inherent frequency nonlinearity of an STO leads to distortions and significant broadening of the spectral line because amplitude fluctuations are expected to create additional phase noise.[7–9] In order to clarify this issue, a detailed experimental study of the temporal coherence and the related linewidth is necessary. So far, attempts to measure the nonlinear coefficients has been restricted to frequency domain spectral measurements, despite the greater performance of time-domain techniques.[10–14]

In this work we report time domain measurements of the microwave voltage noise related to a spin-torque induced oscillation of the synthetic antiferromagnetic (SAF) layer of a MgO MTJ nanopillar.[15] We observe that the signal is affected by both phase and amplitude noise. By constructing the analytical signal using a Hilbert transform of the measured signal, we extract the time-varying power and phase. We demonstrate the correlation between those two and extract the nonlinear coupling coefficient as a function of voltage in the framework of spin-wave theory. We also deduce the power restoration rate, $\Gamma_p$, and the linear linewidth, $\Delta\omega_0$, from the autocorrelation function of the power fluctuations and the phase variance, respectively.

The reported measurements were performed on a MTJ with a low resistance-area (RA) product (0.9 $\Omega$ $\mu$m). Similar oscillator behavior was observed on several devices. The stack composition is $Co_{60}Fe_{20}B_{20}$ (3, free layer)/Mg(1.3)[nat. ox.]/$Co_{60}Fe_{20}B_{20}$ (2, reference layer)/Ru(0.8)/$Co_{70}Fe_{30}$ (2, pinned layer)/PtMn(20), thicknesses, given in parentheses, are in nanometer. The deposited multilayer films were then shaped into well-defined rectangular nanopillars (see Ref. 16 for details). The electrical resistance of the junction at room temperature is 267 $\Omega$ and 309 $\Omega$ for the parallel (P) and antiparallel (AP) configurations, respectively, corresponding to a tunneling magnetoresistance ratio of 16%. For applied static magnetic field between 0 and $-85$ mT along the easy axis in the AP state, we observe an oscillation mode with a clear voltage threshold behavior at 296 mV indicating a spin torque induced mode. In the study presented here the applied field is $-32$ mT. The power spectrum is dominated by a peak at 11.1 GHz and the linewidth minimum is 23.5 MHz as a function of voltage. This mode is generated for positive bias voltages when electrons are flowing from the free to the SAF layer, also favoring the AP state. Similar modes on the same devices have been demonstrated to be edge modes of the SAF layer.[15,17]

The time domain traces were recorded with a single-shot oscilloscope of 18 GHz electrical bandwidth. The traces are 20 $\mu$s long with a sampling frequency of 60 giga-samples per second. The time resolution between two consecutive points is 17 ps. We present the data for five values above threshold of applied dc voltage from 310 to 350 mV. For each voltage bias we recorded several traces consecutively. In order to get rid of the low frequency noise, related harmonics, and the background noise, we mathematically applied a 400 MHz band-pass filter centered at 11.1 GHz.

To extract the instantaneous parameters of a time domain signal, we define a complex variable $x_a(t)$, whose real part is the measured signal $v(t)$ and imaginary part is the Hilbert transform (HT) of $v(t)$, i.e., a signal in quadrature with $v(t)$,

[a)]Electronic mail: laurence.bianchini@ief.u-psud.fr.
[b)]Electronic mail: joo-von.kim@u-psud.fr.

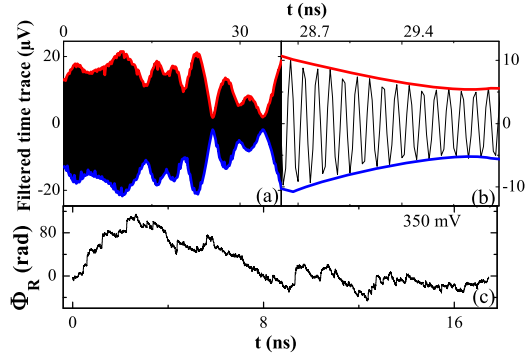

FIG. 1. (Color online) (a) Time-resolved voltage trace at V=350 mV and H=−32 mT in a 37 ns window with the time-varying amplitude $A(t)$ of $x_a(t)$ and $-A(t)$ as the upper and lower envelope traces. (b) A 2.6 ns window zoom of the previous trace. (c) Deduced time-drift of the phase deviation $\Phi_R(t)$ [see Eq. (1)].

$$x_a(t) = v(t) + j\mathbf{HT}[v(t)] \equiv A(t)e^{\Phi(t)}. \quad (1)$$

Hence the modulus and the argument of $x_a(t)$ are the time-varying amplitude $A(t)$ and phase $\Phi(t)$ of the signal $v(t)$, respectively. The time-varying power, is the square of the amplitude and the frequency is the time derivative of the phase.

Figure 1(a) shows an example of an experimental oscillatory time trace after filtering in the above experimental conditions. We observe large amplitude fluctuations [see zoom in Fig. 1(b)]. We checked that the filtering is large enough to avoid significant effects on the amplitude. We see that $A(t)$ and $-A(t)$ [see curves in Figs. 1(a) and 1(b)] are envelopes of the signal. The envelope values are distributed as a Gaussian centered about the mean value. Only the center of the distribution depends on the voltage, as does the rms amplitude of the signal. The distribution width is constant versus voltage.

Let $\delta p(t)$ represent the power fluctuations about the mean value $p(t)=p_0+\delta p(t)$. From Ref. 7 the autocorrelation function of the power fluctuations is given by

$$\kappa_p(\tau) = \langle \delta p(\tau) \delta p(0) \rangle = \mathcal{A}(p_0,\Gamma_p)e^{-2\Gamma_p|\tau|}, \quad (2)$$

where $\Gamma_p$ is the power restoration rate, i.e., it describes how fast power deviations return to the mean value $p_0$. $\Gamma_p$ is expected to increase with increasing voltage as the energy loss is damped out by the spin torque. The stationary power, $p_0$, increases with the applied voltage. For simplicity we chose to fit the decay rate $-2\Gamma_p$ rather than the amplitude $\mathcal{A}(p_0,\Gamma_p)$. The autocorrelation function of the power fluctuations calculated from the experimental data, shown in Fig. 2(a), decays along time, with bumps that are artifacts from band-pass filtering. The filter effects are found not to influence the overall decay rate and can be neglected. The extracted values for $\Gamma_p$ versus voltage are shown in Fig. 2(c) in megahertz. As expected $\Gamma_p$ increases with increasing oscillation power. Note that the extracted $\Gamma_p$ is the inverse of twice the extracted characteristic time and is obtained in rad s$^{-1}$. The relaxation time of the power fluctuations $\tau_p$ that we measure decreases with increasing voltage from 53 ns at 310 mV to 5.4 ns at 350 mV; those being the mean values of the results for several traces. This is consistent with the oscillator theory because the voltage stabilizes the steady state precession.

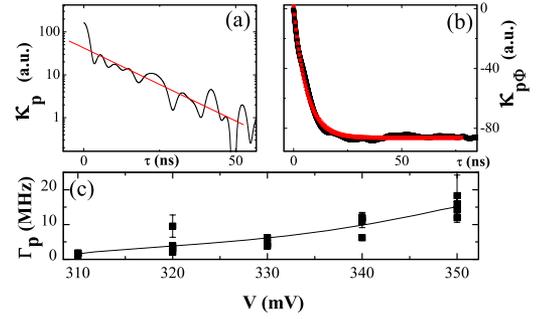

FIG. 2. (Color online) (a) Linear fit of the power autocorrelation function $\kappa_p(\tau)$ on a logarithmic scale at 350 mV. (b) Exponential fit of the cross-correlation function $\kappa_{p\Phi}(\tau)$ between the power and the phase at 350 mV. (c) Plot of the power restoration rate $\Gamma_p$ vs voltage. The line indicates the mean value for each voltage.

From these time domain measurements, it is straightforward to show that the power and the phase are correlated. In Fig. 2(b) we plot the cross correlation product of the power and phase, $\kappa_{p\Phi}(\tau)$, for a bias voltage of 350 mV. It is expected from theory that the power and the phase share the same restoring behavior because of the coupling between the frequency and the oscillation power,[7,9]

$$\omega(t) = \omega(p_0) + \nu\frac{\Gamma_p}{p_0}\delta p(t), \quad (3)$$

$\nu$ is the dimensionless nonlinear frequency coefficient (see Eq. 7 in Ref. 7) and $\omega(p_0)=\omega_0$ is the oscillator nominal frequency. Therefore the only nonzero parameter of $\kappa_{p\Phi}(\tau)$ is proportional to the autocorrelation of the power fluctuations $\kappa_p(\tau)$. The experimental $\kappa_{p\Phi}(\tau)$ decays exponentially [Fig. 2(b)]. We obtain the same characteristic time for $\tau_{p\Phi}$ than the value of $\tau_p$ extracted for a given voltage. The deviation between $\tau_p$ and $\tau_{p\Phi}$ is 9%. While such similar coherence times demonstrate the existence of phase-amplitude coupling, they do not give any information about its strength. It is then necessary to quantify the phase fluctuations.

From the Hilbert transform method we only know the phase between $[-\pi,+\pi]$. Figure 1(c) shows this converted into phase deviation, $\Phi_R(t)$, from its expected phase advance (nominal frequency) accumulated along 17 ns,

$$\Delta\Phi(t) = \int^t \omega(t')dt' = \omega_0 t + \Phi_R(t). \quad (4)$$

We observe that, unlike amplitude noise, the phase drifts along time as a random walk. $\Phi_R(t)$ is the phase fluctuation, and the mean value of the phase noise is nonzero for finite time scale because no restoring force acts on it. As a result, any perturbation of the oscillator causes the phase to drift, which explains why the phase noise is the dominant mechanism for spectral line broadening of any oscillators. Moreover, the consequences of phase noise are drastic for the coherence of the oscillations. The greater the phase noise, the broader the linewidth becomes because the phase noise does not affect the total power in the signal, it only affects its distribution about the central frequency.

The phase variance is used to study the phase dispersion

$$\Delta\Phi_R^2(t) = \langle \Phi_R^2(t) \rangle - \langle \Phi_R(t) \rangle^2. \quad (5)$$

To take into account that the mean value of the phase fluctuates very much over finite time, we cut the measurement into segments and sum the dispersion of the phase of each segment averaged over its proper mean value. From Ref. 7 the phase variance is expected to have the form

$$\Delta\Phi_R^2(t) = 2\Delta\omega_0\left[(1+\nu^2)|t| - \nu^2 \frac{1-e^{-2\Gamma_p|t|}}{2\Gamma_p}\right]. \quad (6)$$

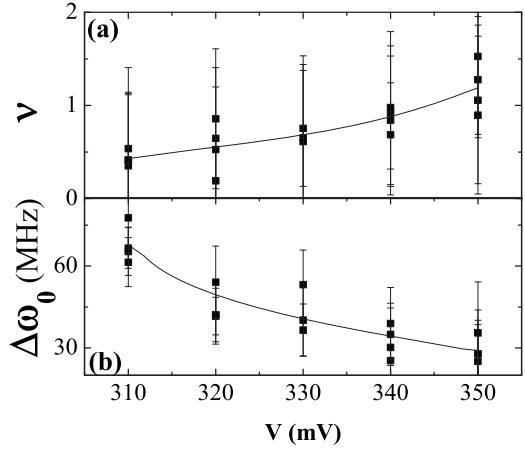

FIG. 3. (a) The nonlinear coefficient $\nu$ and (b) the linear linewidth $\Delta\omega_0$ as a function of applied bias voltage. Lines indicate the mean values for each voltage.

$\Delta\omega_0$ is the linear linewidth. Because the STO is nonlinear, the variance has a nonlinear dependence on the time. On short time scales the variance is quadratic whereas for time intervals much larger than the characteristic time $1/\Gamma_p$ only, the variance starts to grow linearly with time. The nonlinear term in the expression of the variance is one possible origin of a non-Lorentzian line shape. We fit the calculated phase variance from the time series data with the full analytical form above, for different averaging windows. From the mean of the values distribution, where the error is found to be minimum, we extract the most probable $\nu$ and $\Delta\omega_0$.

From the theory we get that the frequency nonlinearity broadens the linewidth by a factor of $(1+\nu^2)$, see Eq. (6). Hence, it is crucial to know the nonlinear coefficient $\nu$ to understand the nature of the linewidth. In Fig. 3(a) we present the extracted coefficient, $\nu$, for five applied voltages above threshold. For each voltage we present $\nu$ from several time traces showing a small dispersion of the values. We find that $\nu$ is comprised between 0 and 1 and slightly increases with increasing voltage. Note that the variation with the voltage is the same order of magnitude than the error bars. For oscillators of most other technologies, $\nu \ll 1$ and the nonlinearity is negligible. STOs differ from conventional oscillators because their nonlinearity is substantial and affects many of their properties. It was predicted that the value of $\nu$ can be both positive or negative, resulting in blueshift or redshift of the frequency and its value depends on the applied field magnitude and orientation.[10,18] One previous study of the nonlinear coefficient extracted from the linewidth dependence on the inverse power[10] found such small values for $\nu$ between 2.5 and 3 for a free layer mode with similar linewidth. For the SAF mode studied here, the frequency only slightly depends on the voltage, hence $\nu$ was expected to be small.

From the fits of the phase variance linear part we also extract $\Delta\omega_0$, the linear linewidth, as a function of voltage, shown in Fig. 3(b). We find values of $\Delta\omega_0$ that are the same than that of the spectral linewidth measured with a spectrum analyzer, which is consistent with the small value of $\nu$ found. $\Delta\omega_0$ decreases with the applied voltage because it results from the competition between the thermal energy $k_BT$ and the oscillation energy.[7–9]

In conclusion, we have provided a direct experimental proof of phase-amplitude coupling in STOs. The time-domain analysis, we have developed, allows phase and amplitude fluctuations to be quantified, which should also prove useful for studying the role of nonlinearity in other systems (such as vortex-based oscillators).

We thank Singulus Technologies A.G. for the layer deposition in a Timaris PVD system. This work was supported by the Triangle de la Physique under Contract No. 2007-051T.